# Sunspot Group Numbers Since 1900 and Implications for the Long-term Record of Solar Activity

Leif Svalgaard[1] and Kenneth H. Schatten[2]
Submitted May 2017

**Abstract**

Recent work on improving and revising estimates of solar activity [Clette et al., 2014] has resulted in renewed interest in what has been called the longest running 'Science Experiment'. We compare four reconstructions of solar activity as reflected in the number of sunspot groups ('active regions') constructed by different authors using very different methods. We concentrate on the period since AD 1900 where the underlying solar and geomagnetic data are plentiful and of sufficient quality and find that all four methods yield essentially the same Sunspot Group Number series. We take that as indicating that protracted and pernicious criticisms of the individual methods are neither fruitful nor helpful and we suggest that future efforts be directed towards understanding the specific reasons why the methods give discordant results for centuries prior to the 20th. The main area of disagreement occurs during the last 25 years of the 19th century and feeds back into the time prior to that. The solar Extreme Ultraviolet flux can be reconstructed since the 1740s [Svalgaard, 2016] and with suitable scaling fits the Svalgaard & Schatten [2016] Sunspot Group Number series since 1865 very well, so we argue that the discordant group series have problems once we move out of the 20th century, and that the community should concentrate on finding out what those are, so a true and useful consensus can emerge.

## 1. Introduction

For a hundred years after Rudolf Wolf's death, his celebrated Relative Sunspot Number (which we shall refer to as the Wolf Number[3], *W*) reigned supreme as *the* measure of solar activity. Wolf was able to extend the *W* series back to AD 1700 for yearly averages and to 1749 for monthly averages. Daily values go back to 1818 (with some gaps during the early years). Then in 1994, Hoyt et al. [1994] asked "Do we have the correct reconstruction of solar activity?" and proposed to answer in the negative by constructing the Group Sunspot Number [Hoyt & Schatten, 1998; hereafter HS], based on counting only the number of sunspot groups, arguing that the number of individual, small spots was not well-determined in the early data. A decade ago Svalgaard [2007] quantified a discrepancy between the *W* series and the HS reconstruction and pointed out that a re-assessment was needed. Recently, Svalgaard & Schatten [2016; hereafter SS] finally revisited the issue and brought the Group Number series up-to-date. Their approach has been severely (e.g. Lockwood et al. [2016]), but unsatisfactorily (see Svalgaard & Schatten [2017]), criticized on procedural grounds. Namely, that our methods were "unsound" and "generally invalid" and that our assumptions were leading to "considerable errors in the long-term". Two new articles (Willamo et al. [2017]; hereafter WEA. Chatzistergos et al. [2017]; hereafter CEA) have joined the ever-growing list of 'new and improved' reconstructions. What is truly remarkable, though, is that since the year 1900 where the underlying data are plentiful and of good quality, all these reconstructions agree within a few percent, regardless of methodology and of claims of being superior to all the others; a conclusion also reached by Cliver [2016]. This means that in spite of the criticisms, all the methods are equally satisfactory, provided the data are good and in particular that the 'standard' or 'reference' observer records are stable and free of anomalies. In this article we quantify the astounding degree of agreement obtained in spite of the very diverse methods and choices made by HS, SS, CEA, and WEA and compare this new consensus for the 20$^{th}$ century with other solar indices with equally impressive agreements.

---

[1] Stanford University, Stanford, CA 94305, USA
[2] a.i. solutions, Lanham, MD 20706, USA
[3] « on pourrait la nommer *Série de R. Wolf*. On pourrait se moquer de cette prétention; mais puisqu'il existe des auteurs sans conscience on est forcé de défendre sa propriété », Wolf [1877].

**2. Building the Composite**

The four articles (HS, SS, CEA, and WEA) all include annual values for the group counts. Because of slightly different choices of reference observers and of secondary observers, the counts are not exactly identical, although nearly so. To compare the series we (arbitrarily) regress the annual values of each series to the SS values, Figure 1.

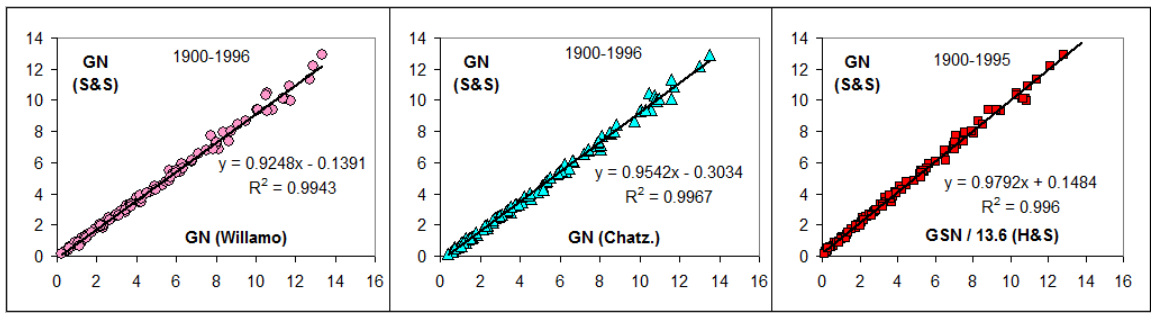

**Figure 1.** Panels from left to right for WEA, CEA, and HS yield linear correlations with SS (with slope and offset) with coefficients of determination $R^2$ in excess of 0.994 and offsets numerically smaller than 1/3 group, covering almost all of the 20$^{th}$ century (1900-1996). The HS group sunspot numbers (GSN) were originally scaled to match the $W$ number. For the purposes of the present article we have de-scaled the HS data back to group counts in order to use the same units.

We first note that the regression lines are straight, i.e. that the correlations are linear. The small offsets are different, close to zero, and therefore probably unphysical (and in any case have no influence on the size of the sunspot cycles measured by the count at maximum), but we keep them as is in the scaling towards the SS counts in the next step, simply because they make no difference for the final result. Figure 2 shows the result of applying the regression relations found on the Figures to each series to put them on the SS scale and then computing an average composite series.

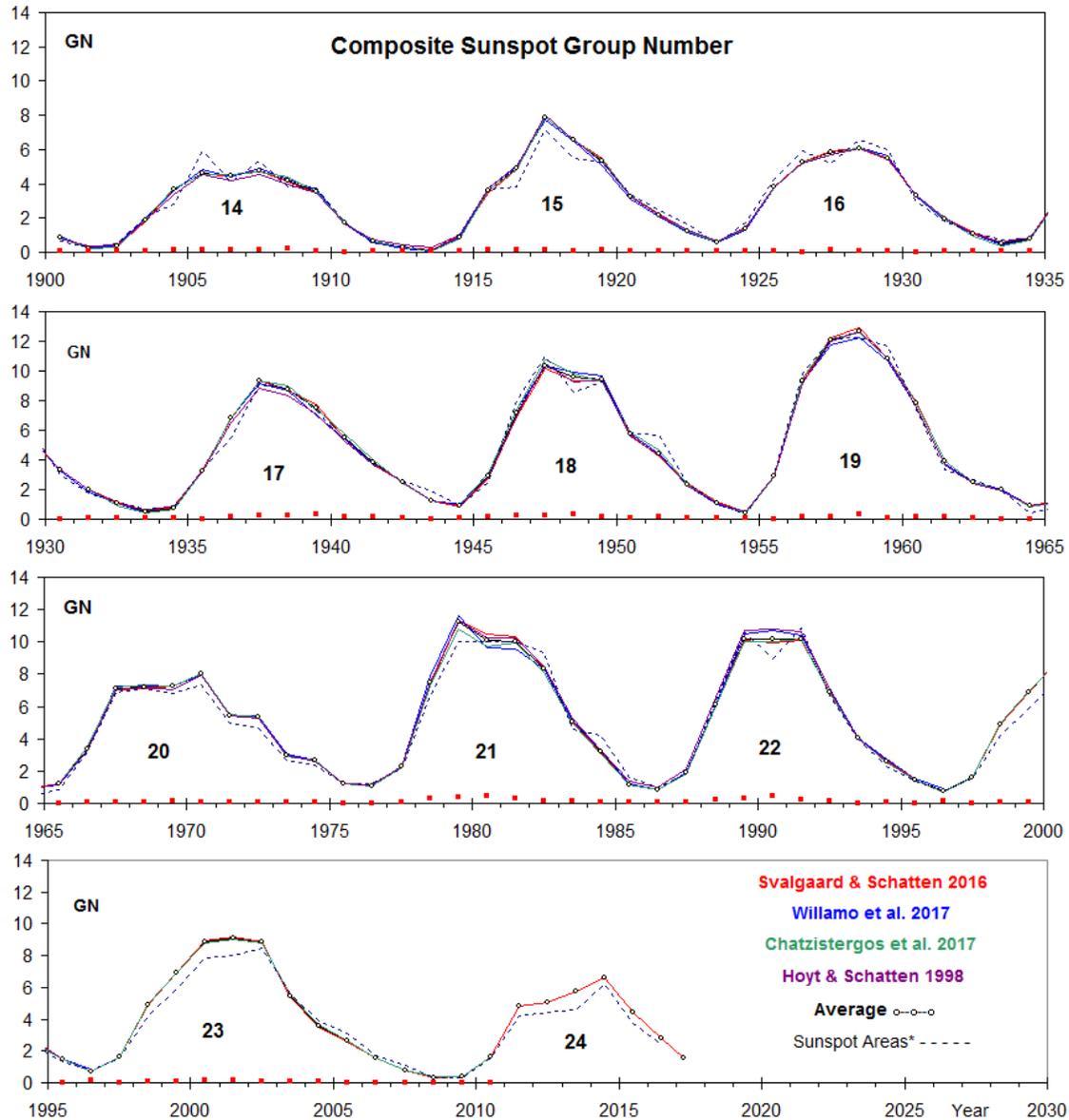

**Figure 2.** Annual values of the scaled reconstructions from the set of four reconstructions we are considering, plotted with different colors as per the legend in the lower panel. The curves are so close that they are hard to distinguish individually. Red squares at the bottom of each panel show the standard deviation of the values that went into the average of the curves which is marked by a curve with small circles. The dashed line shows the reconstruction using the observed (i.e. foreshortened) average sunspot areas for each year. See the text in section 3 for a discussion of the sunspot areas.

Since it is so hard to see the differences between the reconstructions, we show in Figure 3 scatter plots of the individual reconstructions versus the composite average. It is rare to find fits better than $R^2 = 0.994$ for almost one hundred data points, so we consider the resulting composite to be an *excellent* representation of the observed activity as determined by the sunspot group numbers. How useful it is, depends on how it compares with other, perhaps more physical, solar indices and activity measures. This we address in the following sections.

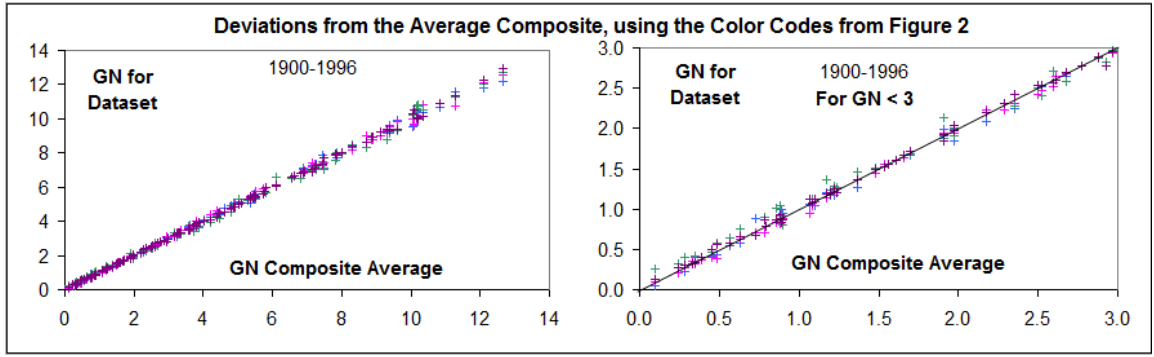

**Figure 3.** Annual values of the scaled reconstructions of the Sunspot Group Numbers from the set of four reconstructions (HS, SS, WEA, CEA) we are considering, plotted against the Composite Series. The right panel shows the magnified lower part (GN < 3) of the comparison.

## 3. Comparison with Sunspot Areas

The long series of sunspot areas published in the Photoheliographic Results from the Royal Greenwich Observatory[4] (RGO) amended by observations from other sources [Balmaceda et al., 2009 and later updates] is often considered to be homogeneous and objective, so it is of considerable interest to compare the areas (Figures 4 and 2, dashed curve) with our Composite series.

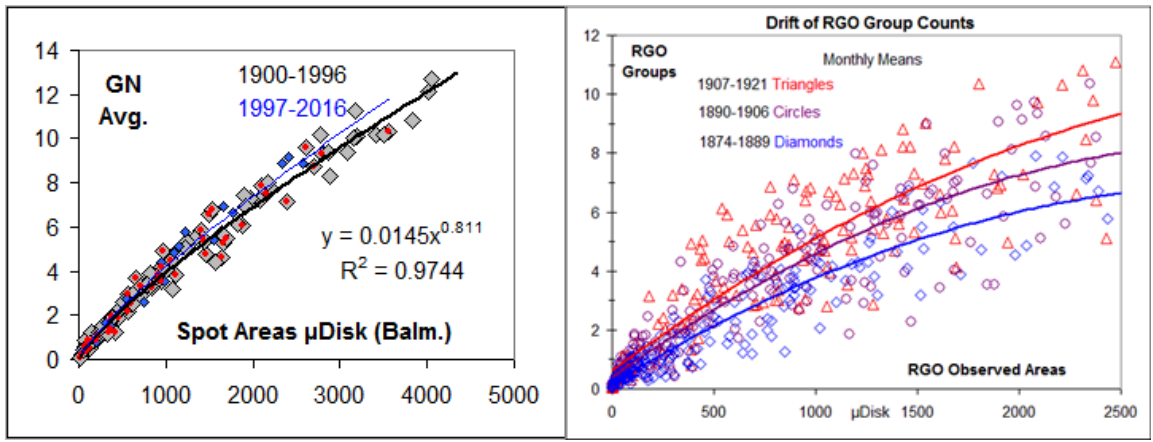

**Figure 4.** (Left) Annual values of the Composite Group Number against the area of the sunspots observed (i.e. foreshortened) on the solar disk in units of millionths of the disk (grey diamonds) for 1900-1996. The data points for the first half of the series are marked with a red dot. Points later than 1996 are plotted separately as small blue diamonds. (Right) Monthly values of the Group count by the RGO observers, showing that their counting was not homogenous before ~1905 (Svalgaard & Schatten [2017]).

---

[4]ftp://ftp.ngdc.noaa.gov/STP/publications/miscellaneous/greenwich_publications/

The dashed curve on Figure 2 (constructed using the power-law regression equation on Figure 4) fits the average Composite very well, *except* for the time after ~1996, where the sunspot area-derived curve falls below the Composite. If this is a real difference, it means that after ~1996 the areas of the spots that make up the visible groups are generally smaller than those of spots observed before 1996. Either the spots are now smaller and/or they are less numerous than before ~1996. We do not here entertain a discussion about whether the time series of sunspot areas is accurate or consistent between various observatories, although the usual problems with harmonization of data from multiple sources cannot be ruled out.

**4. Comparison with Inferred EUV and F10.7**

Svalgaard [2016] showed that the solar extreme ultraviolet flux (EUV) below wavelength λ 102.7 nm can be reliably inferred from the diurnal variation of the geomagnetic field, extending Wolf's and Gautier's original discoveries of 1852[5]. The EUV creates the ionospheric E-region and dynamo action maintains an electric current, the magnetic effect of which is readily observed at the surface (as discovered by Graham in 1722). The (observable) λ 10.7 cm microwave flux (F10.7) is generally thought to be a good proxy for the EUV flux [Tapping, 2013], so it is of interest to compare the inferred EUV flux (given by the range of the diurnal variation) and observed F10.7 flux with our new group number composite, Figure 5.

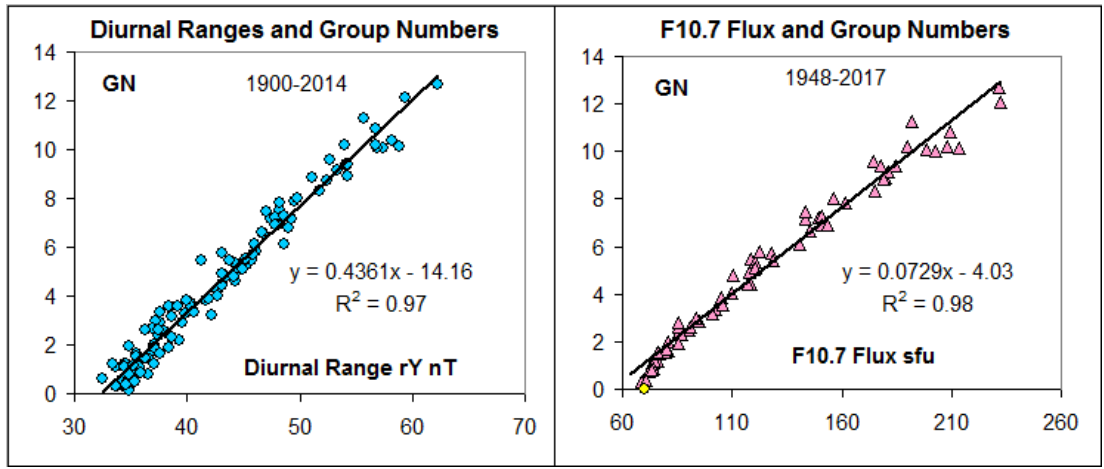

**Figure 5.** (Left) Annual values of the Composite Group Number against the diurnal range, *rY*, of the variation of the East Component of the geomagnetic field at mid-latitudes. (Right) Annual values of the Composite against the F10.7 microwave flux. Generally, $F10.7$ flux = $(rY/4)^2$, [Svalgaard, 2016].

There is a small non-linearity for very low activity, perhaps due to emission from occasional coronal areas above the limb without visible photospheric sunspots, but apart from that, simple linear regression equations 'explain' 97-98% of the observed variation. For solar cycle 24, the F10.7 flux was 70 for days where the actual, modern group

---
[5] "Wer hätte noch vor wenigen Jahren an die Möglichkeit gedact, aus den Sonnenfleckenbeobachtungen ein terrestrisches Phänomen zu berechnen" Wolf [1859]

number was zero (yellow diamond on Figure 5), well above its quiet-sun background of 65 solar flux units (sfu) [Schonfeld et al., 2015].

Reconstructing the Group Number from the Diurnal Range and F10.7 using the regression relations from Figure 5 allows us to calculate a pseudo-group number from the microwave flux and the diurnal ranges.

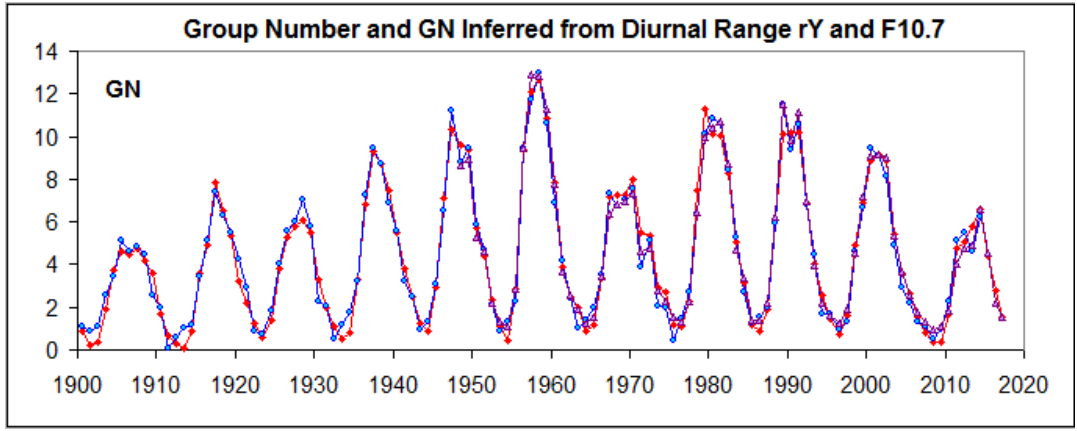

**Figure 6.** Composite Group Number series from solar observations (blue dots) and inferred from the Diurnal Ranges (red diamonds) and from the observed F10.7 flux (purple triangles).

Figure 6 demonstrates that we can reasonably well reconstruct the Group Number from observations of the Diurnal Range which also successfully reproduce the observed F10.7 microwave flux. This reinforces our view that the Group Number is a good, objective measure of solar activity and that any and all of the proposed reconstructions (HS, SS, WEA, and CEA) do a good job in deriving the GN from direct solar observations; in particular, the severely criticized Svalgaard & Schatten [2016] series, rendering moot such criticism and related, unfounded concerns.

## 5. Other Solar Indices

Measurements of the ionized Ca II K-line at λ 393.37 nm are a major resource for long-term studies of solar and stellar activity. They are also critical as a ground-based proxy to model the solar ultraviolet flux variation. Full-disk images of the Sun in Ca II K have been available from various observatories for more than 100 years. Calibration of the old spectroheliograms is a difficult and on-going task, but has recently been successfully carried out for the so-called 1-Å Ca II K emission index, defined as the equivalent width of a 1 Ångström band centered on the K-line core. A Ca II Index composite has been constructed from Kodaikanal, Sacramento Peak, and SOLIS/ISS data [Bertello et al., 2017] and is publicly available from the SOLIS website at http://solis.nso.edu/0/iss/. Regular full-disk Ca II K monitoring programs started in 1915 at the Mount Wilson Observatory and we can scale the MWO data to match the Kodaikanal series. The two Ca II K series in turn can be linearly scaled to the geomagnetic Diurnal Range, $rY$, Figure 7. It is evident that the $rY$-index accurately matches the Ca II indices. And that therefore a

Group Number composite derived from *rY* also is a good representation of the variation of the solar magnetic field reflected in the Ca II plage data.

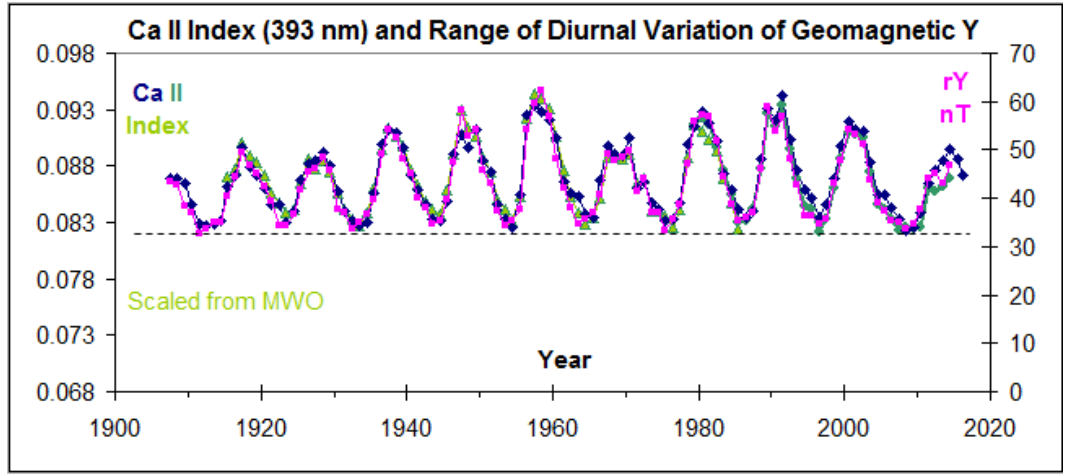

**Figure 7.** Annual values of the Ca II K 1-Å index (blue diamonds) compared to the Diurnal Variation of the East Component of the geomagnetic field (pink squares and right-hand axis) and the MWO data scaled to the Kodaikanal series (green triangles). Note the constant 'floor' at every solar minimum (dashed line).

The "Bremen index" for the Mg II line at λ 280 nm and the total disk-averaged magnetic flux also correlate extremely well with the Diurnal Range and the F10.7 flux, [Svalgaard & Sun, 2016] so we gain considerable confidence in application of geomagnetic and ionospheric data for validating the purely solar reconstructions. We would argue that such 'cross-validation' is necessary to guard against real solar changes masquerading as observer differences, such as what we discuss in the following section.

**6. Decreasing Number of Spots per Group**

The very definition of the Wolf Number, $W = 10\,G + S$ presumes that groups, *G*, are more important than spots, *S*, as Wolf gave groups a weight of 10 (which was about twice the average number of spots per group that he actually counted) and that the number 10 (chosen by Wolf also for convenience), is constant over time. The changing relationship between the number of groups and the area of spots (Section 3) hinted that such constancy may not be taken as a given. Also, there was a change in the relationship between the Wolf Number and the F10.7 flux in about 1990, see e.g. Svalgaard & Hudson [2010], in the sense that too few spots were reported for a given F10.7 flux. Lefèvre & Clette [2011] found that the Sun has shown a deficit in small spots since at least the year 2000. The occurrence rate of the smallest sunspots, and among them the ones with the shortest lifetimes, has more than halved during Cycle 23.

Figure 8 shows that the ratio *W/G* (with Wolfer's counting method) has been about 20 for most of the 20$^{th}$ century, but has been decreasing the last three solar cycles. The average number of spots per group $S/G = W/G - 10$ has then also decreased from about 10 to now only about half of that.

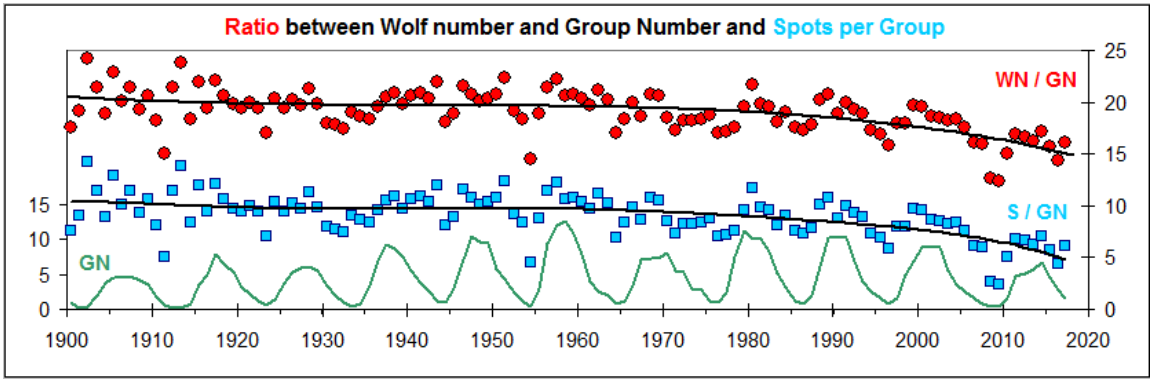

**Figure 8**. The number of spots per group as a function of time (blue squares) and the ratio between the Wolf Number and the Group Number (red dots).

To remove issues with normalization, we investigate this decrease using raw data from the German SONNE network of sunspot observers [SONNE, 2017] and from the long-running Swiss station Locarno [Locarno, 2017], the latter serving as reference or 'pilot' station for the Wolf Number. From each data source, we extract the number, $G$, of groups and the number, $S$, of 'spots' reported by the observers. 'Spots' is in quotation marks because Waldmeier, and to this day Locarno as well [Svalgaard et al., 2017], weighted larger spots more strongly than small spots, so the weighted 'spot' count will be 30-50% larger than the raw count where each spot is counted only once as in Wolf's and Wolfer's original scheme[6]. The SONNE observers do not employ weighting: each spot is counted only once. It is important that for both groups of observers, the counting methods (albeit different) have been unchanged over the period of interest. For both groups, the decrease of $S/G$ is evident, Figure 9, for both SONNE and Locarno and is therefore not due to drifts of calibration or decreasing visual acuity of the primary Locarno observer (Sergio Cortesi, since 1957). If the 'missing 'spots' were large spots with significant magnetic flux, one would expect F10.7 and $rY$ to decrease as well, contrary to the observed trends, so the missing spots must be the smallest spots, as suggested by Lefèvre & Clette [2011]. This also may be a natural explanation for the decline of the Sunspot Number compared to F10.7. We will therefore have to accept that sunspot groups now have significantly fewer and smaller spots than earlier, which is something to worry about when discussing long-term or secular changes before the year 1900.

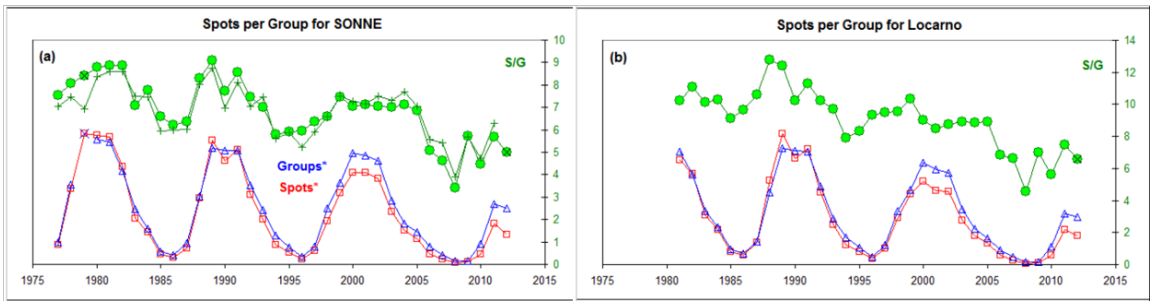

---

[6] Wolfer: "Notiert ein Beobachter mit seinem Instrumente an irgend einem Tage $g$ Fleckengruppen mit insgesamt $f$ Einzelflecken, ohne Rücksicht auf deren Grösse, so ist die daraus abgeleitete Relativzahl jenes Tages $r = k(10g + f)$"

**Figure 9**. (Left) The number of spots per group as a function of time (green circles) for SONNE. The green curve with pluses shows the ratio derived from the raw counts, not corrected with *k*-factors. The lower part of the panel shows the variation of number of groups (blue triangles) and the number of spots (red squares) both scaled to match each other before 1992. Note the decreasing spot count, relative to the group count. (Right) Same, but for Locarno. The ratio is higher because of the weighting of the sunspot count at Locarno, but the downward trend is the same.

## 7. Group Numbers in the 19th Century

Having established that the Diurnal Range is a good proxy for the Sunspot Group Number and that the different techniques for estimating solar activity essentially provide the same solar activity measure for the 20th century, we extend the comparisons of Figure 2 to the 19th Century using the same regression constants and include the scaled Diurnal Range, Figure 10.

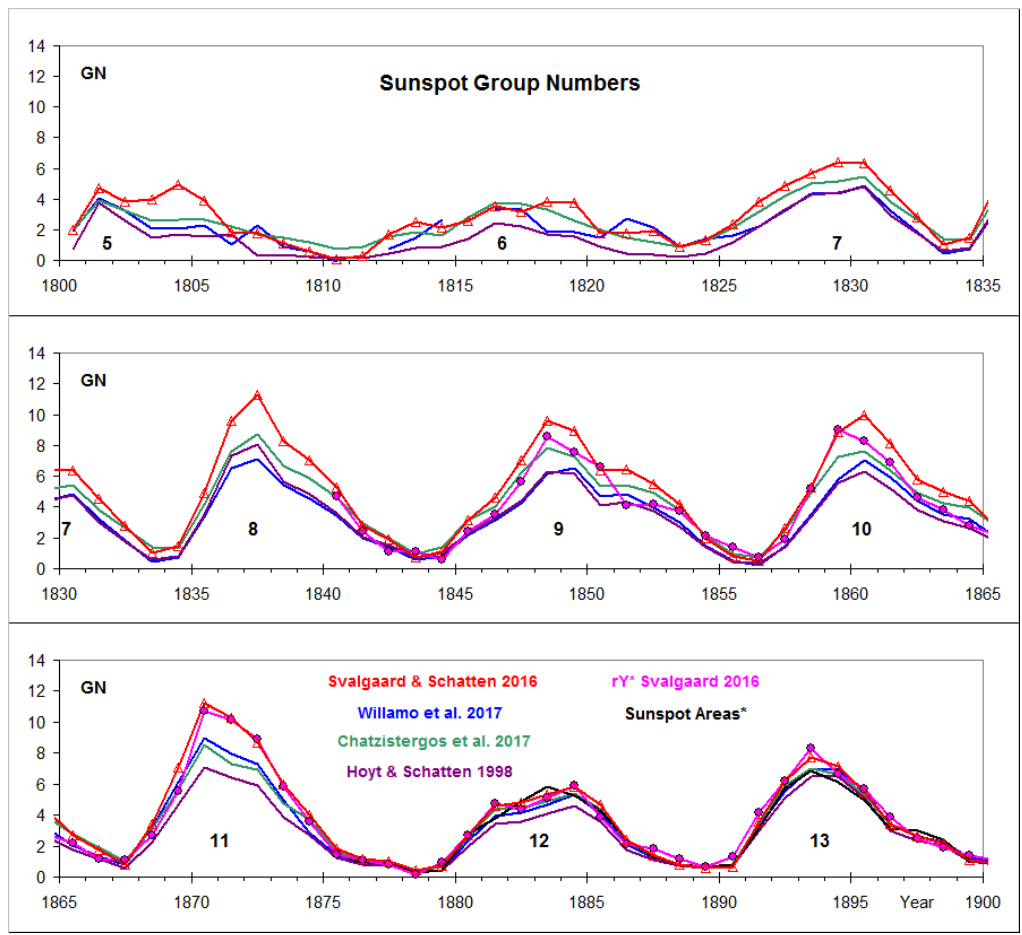

**Figure 10**. (Left) Annual values of the scaled reconstructions from the set of four reconstructions we are considering, plotted with different colors as per the legend in the lower panel. The Diurnal Range, *rY*, scaled according to the regression equation in Figure 5 is shown as the pink curve with dots to compare with the SS reconstruction (red curve with triangles).

The fit of the Diurnal Ranges is a very good match to the SS reconstruction after 1865 and even, although somewhat poorer, before that, possibly reflecting a lesser quality of the data (but still has higher values then than the other reconstructions). Solar Cycle 11 peaking in 1870 is critical for showing that the other reconstructions (HS, WEA, and CEA) fall short during the 19th Century. For earlier cycles, much depends on the quality of the data (both the solar and the geomagnetic data), so to improve the situation we should concentrate on retrieving and digitizing the early data, of which there is plenty [Schering, 1889].

## 8. Group Numbers before the 19th Century

Even before the 'Magnetic Crusade' of the 1840s we have scattered observations of the diurnal variation of the Declination [Svalgaard, 2016; Schering, 1889]. Here we shall limit ourselves to briefly mention the early data mainly collected and published by Wolf and reduced by Loomis [1870, 1873] to the common scale of the magnetic observatory at Prague, Figure 11. Solar Cycles 4, 8, and 11 all appear equally high and on par with cycles 17, 21, and 22.

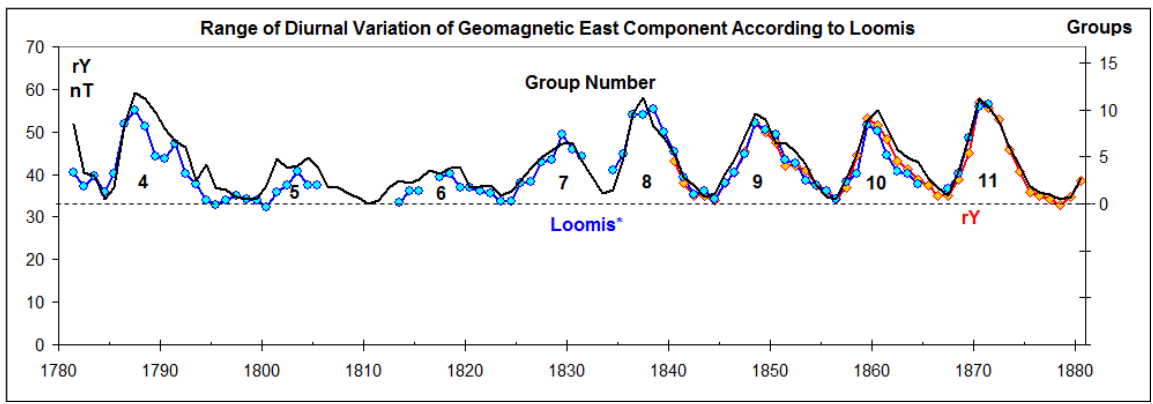

**Figure 11**. The range, $rY$, of the diurnal variation of the geomagnetic East Component determined from the daily range of Declination (Loomis, 1870, 1873) converted to force units in the East direction (blue curve) and then scaled to match $rY$ (red curve, Svalgaard [2016]). The Sunspot Group Number (Svalgaard & Schatten, 2016) is shown (black curve without symbols) for comparison scaled (right-hand scale) to match $rY$.

Loomis drew two important and prescient conclusions: 1) the basal part of the "diurnal inequality (read: variation), amounting at Prague to six [arc] minutes is independent of the changes in the sun's surface from year to year", and 2) "the excess of the diurnal inequality above six minutes as observed at Prague, is almost exactly proportional to the amount of spotted surface upon the sun, and may therefore be inferred to be produced by this disturbance of the sun's surface, or both disturbances may be ascribed to a common cause". We have now observed this 'floor' from the 1780s to the present and it seems possible (even likely) that it is there permanently (see also Schrijver et al. [2011]), with obvious implications for estimating long-term solar activity (e.g. that solar magnetism does not disappear during Grand Minima).

## 9. Conclusion

We have shown that four reconstructions (HS, SS, CEA, and WEA) of the Sunspot Group Number, in spite of very different techniques yield essentially the same measures of solar activity since AD 1900 when the underlying solar and geomagnetic data were plentiful and of good quality. This establishes that all four methods are inherently satisfactory, providing the data are good enough. The reconstructions differ significantly in the centuries prior to the 20$^{th}$ with most of the differences originating during the last 25 years of the 19$^{th}$ century, suggesting that future research be directed at this crucial period. The Svalgaard & Schatten [2016, 2017] backbone method yields a series that agrees best with the geomagnetic evidence. The data used and the analysis can be freely downloaded from http://www.leif.org/research/gn-data.htm.

## Acknowledgements

We are indebted to the following persons and institutions for supplying (hard to get) raw data: Specola Solare Ticinese, Locarno, Switzerland; SILSO, Royal Observatory of Belgium; Prof. Roland Hedewig, Kassel, Germany; Andreas Bulling, Pfullingen, Germany; L.S. thanks Stanford University for support.

## Disclosure of Potential Conflicts of Interest

The authors declare that they have no conflicts of interest.

## References

Balmaceda, L. A., Solanki, S. K., Krivova, N. A., and Foster, S.: A homogeneous database of sunspot areas covering more than 130 years, *J. Geophys. Res.* **114**(A7), CiteID A07104, doi:10.1029/2009JA014299, 2009.

Bertello, L., Marble, A. R., and Pevtsov, A. A.: Ca II K 1-Å Emission Index Composites, National Solar Obs. *Tech. Rep. NSO/NISP-2017-001*, (https://arxiv.org/pdf/1702.00838.pdf), 2017.

Chatzistergos, T., Usoskin I. G., Kovaltsov, G. A., Krivova, N. A., and Solanki, S. K.: New reconstruction of the sunspot group number since 1739 using direct calibration and 'backbone' methods, *Astron. & Astrophys.* in press, (https://arxiv.org/pdf/1702.06183.pdf), 2017.

Clette, F., Svalgaard, L., Vaquero, J. M., and Cliver, E. W.: Revisiting the Sunspot Number - A 400-Year Perspective on the Solar Cycle, *Space Sci. Rev.* **186**, 35, doi:10.1007/s11214-014-0074-2, 2014.

Cliver, E. W.: Comparison of New and Old Sunspot Number Time Series, *Solar Phys.* **291**(9-10), 2891, doi:10.1007/s11207-016-0929-7, 2016.

Hoyt D. V., Schatten K. H., and Nesmes-Ribes, E.: The one hundredth year of Rudolf Wolf's death: Do we have the correct reconstruction of solar activity? *Geophys. Res. Lett.* **21**(18), 2067, doi:10.1029/94GL01698, 1994.

Hoyt, D. V. and Schatten, K. H.: Group Sunspot Numbers: A New Solar Activity Reconstruction, *Solar Phys.* **181**(2), 491, 1998.

Lefèvre, L., Clette, F.: A global small sunspot deficit at the base of the index anomalies of solar cycle 23, *Astron. & Astrophys.* **536**, id.L11, doi:10.1051/0004-6361/201118034, 2011.

Locarno: Specola Solare Ticinese, *Drawings Archive*, http://www.specola.ch/e/drawings.html, 2017.

Loomis, E.: Comparison of the mean daily range of Magnetic Declination, with the number of Auroras observed each year, and the extent of the black Spots on the surface of the Sun, *Am. Journ. Sci. Arts*, 2nd Series **50**(149), 153, 1870.

Loomis, E.: Comparison of the mean daily range of the Magnetic Declination and the number of Auroras observed each year, *Am. Journ. Sci. Arts*, 3rd Series **5**(28), 245, 1873.

Schering, K.: Die Entwicklung und der gegenwartige Standpunkt der erdmagnetische Forschung, *Geograph. Jahrbuch* **13**, 171, http://www.leif.org/research/Schering-1889.pdf, 1889.

Schonfeld. S. J., White, S. M., Henney, C. J., Arge, C. N., and McAteer, R. T. J.: Coronal sources of the solar $F_{10.7}$ radio flux, *Astrophys. Journ.* **808**(29), 1, doi:10.1088/0004-637X/808/1/29, 2015.

Schrijver, C. J., Livingston, W. C., Woods, T. N., and Mewaldt, R. A.: The minimal solar activity in 2008–2009 and its implications for long-term climate modeling, *Geophys. Res. Lett.* **38**, L06701, doi:10.1029/2011GL046658, 2011.

SONNE: *Results*, Sunspots 1977-2017, http://www.vds-sonne.de/index.php, 2017.

Svalgaard, L.: Calibrating the Sunspot Number using "the Magnetic Needle", *CAWSES News* **4**(1), 6, http://www.bu.edu/cawses/calbrating_sunspot_number_using_mag_needle.pdf, 2007.

Svalgaard, L. and Hudson, H. S.: The Solar Microwave Flux and the Sunspot Number, in SOHO-23: Understanding a Peculiar Solar Minimum, *ASP Conference Series* **428**, 325, S. R. Cranmer, J. T. Hoeksema, and J. L. Kohl, eds., 2010.

Svalgaard, L.: Reconstruction of Solar Extreme Ultraviolet Flux 1740-2015, *Solar Phys.* **291**(10), 2981, doi:10.1007/s11207-016-0921-2, 2016.

Svalgaard, L., and Sun, X: Variation of EUV matches that of the Solar Magnetic Field and the Implication for Climate Research, *LWS-SDO 2016 Meeting*, Burlington, Vermont, http://www.leif.org/research/EUV-Magnetic-Field.pdf, 2016.

Svalgaard, L., Cagnotti, M., and Cortesi, S.: The Effect of Sunspot Weighting, *Solar Phys.* **292**(2), article id.34, doi:10.1007/s11207-016-1024-9, 2017.

Svalgaard, L. and Schatten, K. H.: On the Sunspot Group Number Reconstruction: The Backbone Method, *Solar Phys.* (submitted), arXiv:1704.07061 [astro-ph.SR], 2017.

Tapping, K. F.: The 10.7 cm solar radio flux (F10.7), *Space Weather* **11**, 394, doi:10.1002/swe.20064, 2013.

Willamo, T., Usoskin, I. G., and Kovaltsov, G. A.: Updated sunspot group number reconstruction for 1749–1996 using the active day fraction method, *Astron. & Astrophys.* in press, doi:/10.1051/0004-6361/201629839, 2017.

Wolf, R.: Mittheilungen über die Sonnenflecken, **IX**, 217, 1859. And: Entdeckung des Zusammenhanges zwischen den Declinationsvariationen der Magnetnadel und den Sonnenflecken, *Mitth. der naturforsch. Gesell. Bern 224–264*, Nr. **245**, 1852.

Wolf, R.: Mémoire sur la période commune à la fréquence des taches solaires et à la variation de la déclinaison magnétique, *Memoirs Royal Astr. Soc.* **43**, 199, 1877.